\documentclass[preprint,number]{elsarticle}

\usepackage{graphicx}
\usepackage{amsmath}
\usepackage{url}
\usepackage{epstopdf}  
\usepackage{slashed}

\begin{document}

\title{Constraining P and CP violation in the main decay of the neutral Sigma hyperon}

\author[upps]{Shankar Sunil Nair}
\author[upps]{Elisabetta Perotti}
\author[upps]{Stefan Leupold}

\address[upps]{Institutionen f\"or fysik och astronomi, Uppsala Universitet, Box 516, S-75120 Uppsala, Sweden} 

\begin{abstract}
On general grounds based on quantum field theory the decay amplitude for $\Sigma^0 \to \Lambda \gamma$ consists of a 
parity conserving magnetic and a parity violating electric dipole transition moment. 
Because of the subsequent self-analyzing 
weak decay of the $\Lambda$ hyperon the interference between magnetic and electric dipole transition moment leads to an 
asymmetry in the angular distribution. Comparing the decay distributions for the $\Sigma^0$ hyperon and its antiparticle gives
access to possible C and CP violation. Based on flavor SU(3) symmetry the present upper limit on the 
neutron electric dipole moment can be translated to an 
upper limit for the angular asymmetry. It turns out to be far below any experimental resolution that one can expect in the 
foreseeable future. 
Thus any true observation of a CP violating angular asymmetry would constitute 
physics beyond the standard model, even if extended by a CP violating QCD theta-vacuum-angle term.
\end{abstract}


\maketitle

\section{Introduction}
\label{sec:intro}

There is much more matter than antimatter in the universe. If this is not just by chance but has a dynamical origin, then 
an explanation of this baryon asymmetry should come from the realm of particle physics. 
Based on Sakharov's conditions \cite{Sakharov:1967dj} this has spurred the search for baryon decays and reactions that show CP 
violation \cite{bigi2}. (Here 
C denotes charge conjugation symmetry and P parity symmetry.) Two directions of such searches are 
weak decays of baryons \cite{Holmstrom:2004ar,White:2007zza,Aaij:2016cla} and 
electric dipole moments (EDMs) \cite{Pospelov:2005pr}. 

In the present work we propose to study a reaction that is in between these two types of reactions, the decay of the neutral 
ground-state Sigma hyperon to a photon and a Lambda hyperon, $\Sigma^0 \to \gamma \Lambda$. This electromagnetic baryon 
decay could in principle show an interference between a parity conserving and a parity violating amplitude. The latter would 
come from an electric dipole transition moment (the former from a magnetic transition moment). 
Note that the $\Lambda$ hyperon decays further into pion and proton on account 
of the weak interaction \cite{Olive:2016xmw}. 
The possible interference in the first decay can then be observed as an angular asymmetry in the 
decay products of the, in total, three-body decay $\Sigma^0 \to \gamma \, \pi^- p$.
Comparing the asymmetry parameters for the particle decay, $\Sigma^0 \to \gamma \Lambda$, and for the 
corresponding antiparticle decay, $\bar \Sigma^0 \to \gamma \bar\Lambda$, one can search for C and CP violation.

Let us add right away that the conservative expectation is that one would not find an effect of P, C or CP violation in this 
decay. This will be substantiated by our explicit estimates given below. Yet, if theory predicts that something is very 
small, then it might be worth checking this experimentally. Even if one ``only'' establishes an upper limit, this can help to 
constrain beyond-standard-model developments. Needless to add that if one found an effect not predicted by established theory, 
then this would be sensational. 

Actually already in 1962 the interplay of magnetic and electric dipole transition moment for the $\Sigma^0$ decay has been 
addressed in \cite{Dreitlein:1962zz}, though under the implicit assumption of CP conservation. 
In the present work we will be more 
general. In addition, to the best of our knowledge, the search for such interference effects, albeit suggested so early, 
has never been conducted. With recent and ongoing experiments on hyperon production 
(e.g.\ at ELSA \cite{Jude:2015rpp}, J-LAB \cite{McCracken:2015coa,Paterson:2016vmc}, GSI \cite{Munzer:2017hbl,Lalik:2017xkw}, 
BEPC II \cite{Li:2016tlt,Schonning:2017ctv}, KEKB \cite{Niiyama:2017wpp}, CESR \cite{Dobbs:2017bmt}) and 
in particular with the upcoming PANDA experiment \cite{Lutz:2009ff,Papenbrock:2016bum} as a hyperon-antihyperon factory, 
it is absolutely timely to establish at least upper limits for the angular asymmetry of the $\Sigma^0$ decay.  

The process $\Sigma^0 \to \gamma \Lambda$ constitutes the main decay of the ground state $\Sigma^0$. 
Its life time is governed by this decay \cite{Olive:2016xmw}. 
Therefore the $\Sigma^0$ lives much longer than hadronic resonances that decay on account of the strong interaction, but much
shorter than weakly decaying particles. This makes the analysis of $\Sigma^0$ decays challenging, as 
already pointed out in \cite{Dreitlein:1962zz}. Hopefully these problems can be diminished by the increasing production 
rate of $\Sigma^0$ particles and detector quality. 

Let us relate and compare the process $\Sigma^0 \to \gamma \Lambda$ in more detail to the mentioned searches for CP violation 
in weak baryon decays and for EDMs. In our case there is an interference between two amplitudes, a large and a small one. 
The first one is related to the magnetic transition moment. This amplitude is parity conserving and 
compatible with Quantum Electrodynamics (QED) and Quantum Chromodynamics (QCD).
The second one is related to the electric dipole transition moment and is parity violating. 
Thus one expects small effects 
right away. This is in contrast to similar radiative decays where both amplitudes are caused by the weak interaction, for 
instance $\Xi^0 \to \Lambda \gamma$ \cite{Batley:2010bp}; see also the note on ``Radiative Hyperon Decays'' from the 
Particle Data Group \cite{Olive:2016xmw}. For the weak radiative decays the angular asymmetry is appreciably large, 
but the difference between particle and antiparticle decay is in most cases so far unmeasurably small. Only recently first 
evidence for a CP violating decay of the $\Lambda_b$ baryon has been reported \cite{Aaij:2016cla}. For our case we expect very small effects for the angular asymmetry and for the particle-antiparticle differences. 

Though nature violates all discrete symmetries P, C and CP\footnote{In quantum field theory the third discrete symmetry, 
time reversal symmetry T, is intimately tied to CP via the CPT theorem \cite{Weinberg:1995mt}. In our decay, time reversal 
arguments are not directly applicable because we do not study the inverse formation process $\Lambda \gamma \to \Sigma^0$. 
Therefore we focus on CP instead of T.}, this happens to different degrees and different interactions behave 
differently. Therefore in an analysis of the decay $\Sigma^0 \to \gamma \Lambda$ it is useful to distinguish 
conceptually between a scenario of C violation but CP conservation and a scenario of C conservation but CP violation. We will 
present observables to test both scenarios for the process of interest. 
Yet for our concrete estimates we will focus on strong CP violation.

From an experimental point of view, the strong interaction conserves P, C and CP separately. Yet from the theory side it would 
be natural to expect CP violation in QCD based on the non-trivial topological structure of the non-abelian 
gauge theory \cite{Bailin:1986wt}. Such a ``theta-vacuum-angle term'' is C conserving but P and CP violating. In particular, it gives rise to 
an EDM of the neutron. So far, no EDM of the neutron has been experimentally 
established \cite{Lamoreaux:2009zz}, but the very small upper limit raises the question why CP violation in the strong sector is so 
unnaturally small (the ``strong-CP problem'') \cite{Peccei:1998jt}. Note that the weak CP violation leads to an EDM of the 
neutron that is many orders of magnitude below the experimental upper limit \cite{Pospelov:2005pr}. Therefore we focus on 
possible strong CP violation in the following. 
As we will show below, our decay $\Sigma^0 \to \gamma \Lambda$ is related to the neutron EDM via SU(3) 
flavor symmetry \cite{GellMann:1961ky,Ottnad:2009jw}. Based on the experimental upper limit of the neutron EDM we will provide an upper 
limit for the CP violating effect on the Sigma decay chain.

The rest of the paper is structured in the following way: In the next section we will present a general parametrization 
of the transition moments for radiative decay amplitudes of baryons, discuss the impact of C and CP symmetry/violation 
and relate all this
to the angular asymmetries of baryon and antibaryon decays. 
In section \ref{sec:dec-ang} we will calculate all relevant decay widths and angular distributions. 
In section \ref{sec:results} we will determine 
an upper limit for those effects caused by strong CP violation, which is in turn related to the QCD theta-vacuum-angle term. 
Here baryon chiral 
perturbation theory \cite{Ottnad:2009jw} can be used to relate the neutron EDM to the $\Sigma^0$ decay. Finally a summary 
and a brief outlook will be provided in section \ref{sec:summary}.

\section{Transition moments and angular decay asymmetries}
\label{sec:transmom}

For the coupling of baryons to the electromagnetic current $J_\mu$ we follow in principle \cite{Ottnad:2009jw} but adopt the 
definition of the photon momentum to our decay process:
\begin{eqnarray}
\langle B'(p') \vert J^{\mu} \vert B(p) \rangle = e\, \bar{u}_{B'}(p') \, \Gamma^{\mu}(q) \, u_{B}(p)
\end{eqnarray}
with $q:=p-p'$ and
\begin{eqnarray}
  \Gamma^{\mu}(q) &=& \left(\gamma^{\mu}+\frac{m_{B'}-m_{B}}{q^{2}} \, q^{\mu} \right) F_{1}(q^{2})
   \nonumber  \\ 
  && {}+ i \left(\gamma^{\mu} q^{2}+ (m_{B}+m_{B'}) \, q^{\mu} \right) \, \gamma_{5}\, F_{A}(q^{2})  \nonumber  \\ 
  && {}-\frac{i}{m_{B}+m_{B'}} \, \sigma^{\mu\nu}q_{\nu} \, F_{2}(q^{2}) 
  -\frac{1}{m_{B}+m_{B'}} \, \sigma^{\mu\nu}q_{\nu}\gamma_{5}  \, F_{3}(q^{2})  \,.
\end{eqnarray}
If $B$ and $B'$ have the same intrinsic parity then the functions $F_{1}(q^{2})$ and $F_{2}(q^{2})$ are the P conserving Dirac and Pauli transition form factors. $F_{A}(q^{2})$ and $F_{3}(q^{2})$ are the P violating Lorentz invariant transition form factors and are termed the anapole form factor and the electric dipole form factor, respectively. We note in passing that ideas how to access the $q^2$ dependence of $F_1$ and $F_2$ for the transition of $\Sigma^0$ to $\Lambda$ have been presented in \cite{Granados:2017cib}. 

For the neutron the Pauli form factor at the photon point is related to the anomalous magnetic moment by \cite{Olive:2016xmw}
\begin{eqnarray}
  \label{eq:pauli-neutron}
  F_{2,n}(0) = \kappa_n \approx -1.91
\end{eqnarray}
while the EDM of the neutron is given by
\begin{eqnarray}
  \label{eq:neutron-EDM1}
  d_n = \frac{e}{2 m_n} \, F_{3,n}(0)   \,.
\end{eqnarray}

The decay $\Sigma^0 \to \gamma \Lambda$ is only sensitive to the magnetic (dipole) transition 
moment \cite{Olive:2016xmw,Granados:2017cib}
\begin{eqnarray}
  \label{eq:defmagmoment}
  \kappa_M := F_2(0) \approx 1.98 
\end{eqnarray}
and the electric dipole transition moment
\begin{eqnarray}
  \label{eq:defEDMtrans}
  \kappa_E := F_3(0) \,.
\end{eqnarray}
The (transition) charge must vanish, $F_1(0) = 0$, and the anapole moment $F_A$ does not contribute for real photons, 
technically based on $q^2 =0$ and $q^\mu \epsilon_\mu =0$ where $\epsilon_\mu$ denotes the polarization vector of the photon.
In section \ref{sec:results} we will relate the electromagnetic properties of the neutron and of the $\Sigma$-$\Lambda$ 
transition. 

Following \cite{Olive:2016xmw} the successive decay $\Lambda \to \pi^- p$ is parametrized by the matrix element  
\begin{eqnarray}
  {\cal M}_2 =\bar{u}_{p}(\mathcal{A}-\mathcal{B}\gamma_{5}) \, u_{\Lambda}
  \label{eq:amp2}
\end{eqnarray}
where $\mathcal{A}$ and $\mathcal{B}$ are complex numbers.
To stay in close analogy to this pa\-rametrization we write the decay matrix element for the first decay 
$\Sigma^0 \to \gamma \Lambda$ as 
\begin{eqnarray}
  \label{eq:amp1}
  {\cal M}_1 = \bar u_{\Lambda} \, (a \, \sigma_{\mu\nu} - b \, \sigma_{\mu\nu} \gamma_5 ) \, u_{\Sigma^0} \, 
  (-i) q^\nu \epsilon^{\mu *}  \,.
\end{eqnarray}
The two decay parameters $a$ and $b$ are related to the transition moments via
\begin{eqnarray}
  \label{eq:abkappas}
  a = \frac{e}{m_{\Sigma^0} + m_\Lambda} \, \kappa_M \,, \qquad b = i \, \frac{e}{m_{\Sigma^0} + m_\Lambda} \, \kappa_E  \,.
\end{eqnarray}
The decay asymmetries that will finally show up in the angular distribution of the 
three-body decay $\Sigma^0 \to \gamma \, \pi^- p$ are defined by (see, e.g., \cite{Olive:2016xmw})
\begin{eqnarray}
  \label{eq:alphaSigma}
  \alpha_{\Sigma^0} := \frac{2 {\rm Re}(a^* b)}{\vert a \vert^2 + \vert b \vert^2}
\end{eqnarray}
and
\begin{eqnarray}
  \label{eq:alphaLambda}
  \alpha_{\Lambda} := \frac{2 {\rm Re}(s^* p)}{\vert s \vert^2 + \vert p \vert^2}  \,.
\end{eqnarray}
with $s:=\mathcal{A}$ and $p := \eta \mathcal{B}$. We have introduced $\eta := \vert \vec p_p \vert/(m_p+E_p)$ to compensate for 
the p-wave phase space relative to the s-wave. Here $m_p$ denotes the mass of the 
proton and $E_p$ ($\vec p_p$) its energy (three-momentum) in the rest frame of the $\Lambda$ decaying into pion and proton.

To reveal C and/or CP violation for our processes one has to compare particle and antiparticle decays. To this end we introduce 
the following matrix elements that correspond to \eqref{eq:amp2} and \eqref{eq:amp1}:
\begin{eqnarray}
  {\cal M}_{\bar 2} = -\bar v_{\Lambda} (\bar{\mathcal{A}}-\bar{\mathcal{B}}\gamma_{5}) \, v_{p} \,,
  \label{eq:amp2bar}
\end{eqnarray}
\begin{eqnarray}
  \label{eq:amp1bar}
  {\cal M}_{\bar 1} = -\bar v_{\Sigma^0} \, (\bar a \, \sigma_{\mu\nu} - \bar b \, \sigma_{\mu\nu} \gamma_5 ) \, v_{\Lambda}  \, 
  (-i) q^\nu \epsilon^{\mu *}  \,.
\end{eqnarray}
In analogy to \eqref{eq:alphaSigma} and \eqref{eq:alphaLambda} we also introduce the asymmetries
\begin{eqnarray}
  \label{eq:alphaSigmabar}
  \alpha_{\bar\Sigma^0} := \frac{2 {\rm Re}(\bar a^* \bar b)}{\vert \bar a \vert^2 + \vert \bar b \vert^2}
\end{eqnarray}
and
\begin{eqnarray}
  \label{eq:alphaLambdabar}
  \alpha_{\bar\Lambda} := \frac{2 {\rm Re}(\bar s^* \bar p)}{\vert \bar s \vert^2 + \vert \bar p \vert^2}  \,.
\end{eqnarray}
with $\bar s:=\bar{\mathcal{A}}$ and $\bar p := \eta \bar{\mathcal{B}}$. 

C and CP conservation/violation influence the phases of the parameters $a$ and $b$. 
In general, the products of a decay show some final-state interaction that leads to an additional phase; 
see, e.g., \cite{bigi2} and the note on ``Baryon Decay Parameters'' in \cite{Olive:2016xmw}. It is useful to distinguish these
two effects. To simplify things we note, however, that overall phases can be freely chosen in quantum mechanics. Only
phase differences matter. Consequently we decompose
\begin{eqnarray}
  \label{eq:decomdecFSI}
  b = b_D \, e^{i \delta_F}  \qquad a = a_D
\end{eqnarray}
with the parameters from the direct decay labeled by an index $D$. The phase difference $\delta_F$ between $a$ and $b$ 
encodes the final-state interaction. For this final-state 
interaction between $\Lambda$ and $\gamma$ we assume that it is dictated by (P and C conserving) QED,
which implies that it is the same for $\Lambda$-$\gamma$ and $\bar \Lambda$-$\gamma$. In \ref{sec:appendix} we 
demonstrate that the parity conserving contribution $a$ to the decay amplitude can be chosen real and positive for the 
particle and the antiparticle decay. CP 
conservation would imply that the parity violating contribution $b_D$ is real, while C conservation would imply that $b_D$
is purely imaginary; see \ref{sec:appendix} for details.

The decays $\Lambda \to \pi^- p$ and $\bar\Lambda \to \pi^+ \bar p$ are governed by the weak interaction \cite{Donoghue:1992dd}. 
Here the CP violation is very small. If one ignores CP violation one obtains $\alpha_\Lambda = -\alpha_{\bar\Lambda}$. 
So far the experimental results are
compatible with this assumption of (approximate) CP conservation \cite{Olive:2016xmw}.

\section{Integrated and differential decay widths and angular distributions}
\label{sec:dec-ang}

The matrix element \eqref{eq:amp1} gives rise to the decay width
\begin{eqnarray}
  \label{eq:width-Sigma}
  \Gamma_{\Sigma^0 \to \gamma \Lambda} = (\vert a \vert ^2 + \vert b \vert^2) \, \frac{(m_{\Sigma^0}^2-m_\Lambda^2)^3}{8\pi m_{\Sigma^0}^3} \,.
\end{eqnarray}
For later use we also provide the partial decay width of the $\Lambda$ hyperon:
\begin{eqnarray}
  \Gamma_{\Lambda \to \pi^- p} = \frac{\lambda^{1/2}(m_\Lambda^2,m_p^2,m_\pi^2)}{16\pi m_\Lambda^3} \, R_\Lambda   
  \label{eq:width-Lambda}
\end{eqnarray}
with the K\"all\'en function 
\begin{eqnarray}
  \label{eq:defKaellen}
  \lambda(a,b,c):=a^2+b^2+c^2-2(ab+bc+ac)
\end{eqnarray}
and \cite{Faldt:2013gka}
\begin{eqnarray}
  \label{eq:defRlambda}
  R_\Lambda := \vert \mathcal{A}\vert^2 ((m_\Lambda+m_p)^2-m_\pi^2) + 
    \vert \mathcal{B}\vert^2 ((m_\Lambda-m_p)^2-m_\pi^2)  \,.
\end{eqnarray}
In general, the corresponding decay widths for the antiparticle decays are obtained by replacing the parameters $a$, $b$, 
$\mathcal{A}$, $\mathcal{B}$ by the corresponding parameters for the antiparticles.

\begin{figure}[h]
  \centering
  \begin{minipage}[c]{0.4\textwidth}
    \includegraphics[keepaspectratio,width=\textwidth]{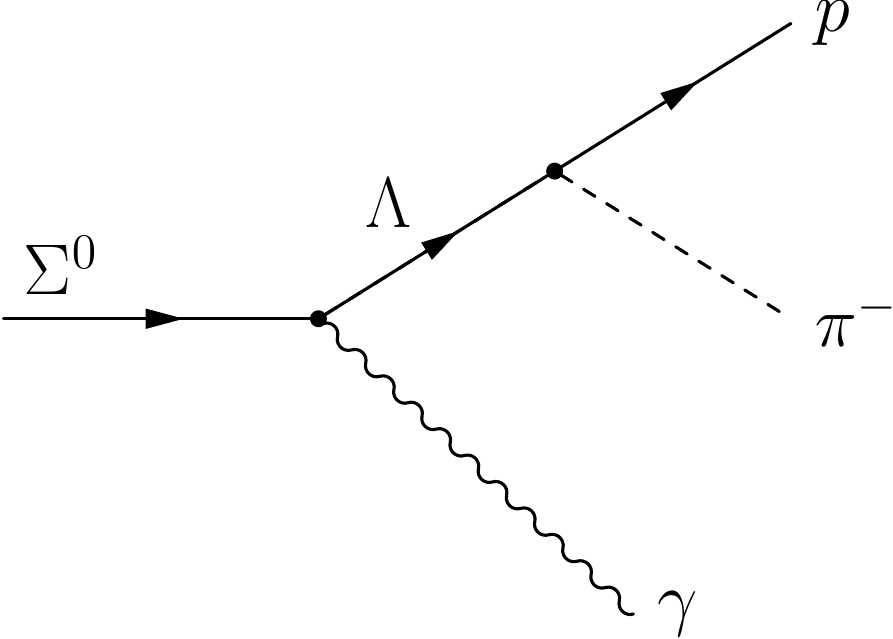}
  \end{minipage}  \hspace*{5em}
  \begin{minipage}[c]{0.3\textwidth}  
    \includegraphics[keepaspectratio,width=\textwidth]{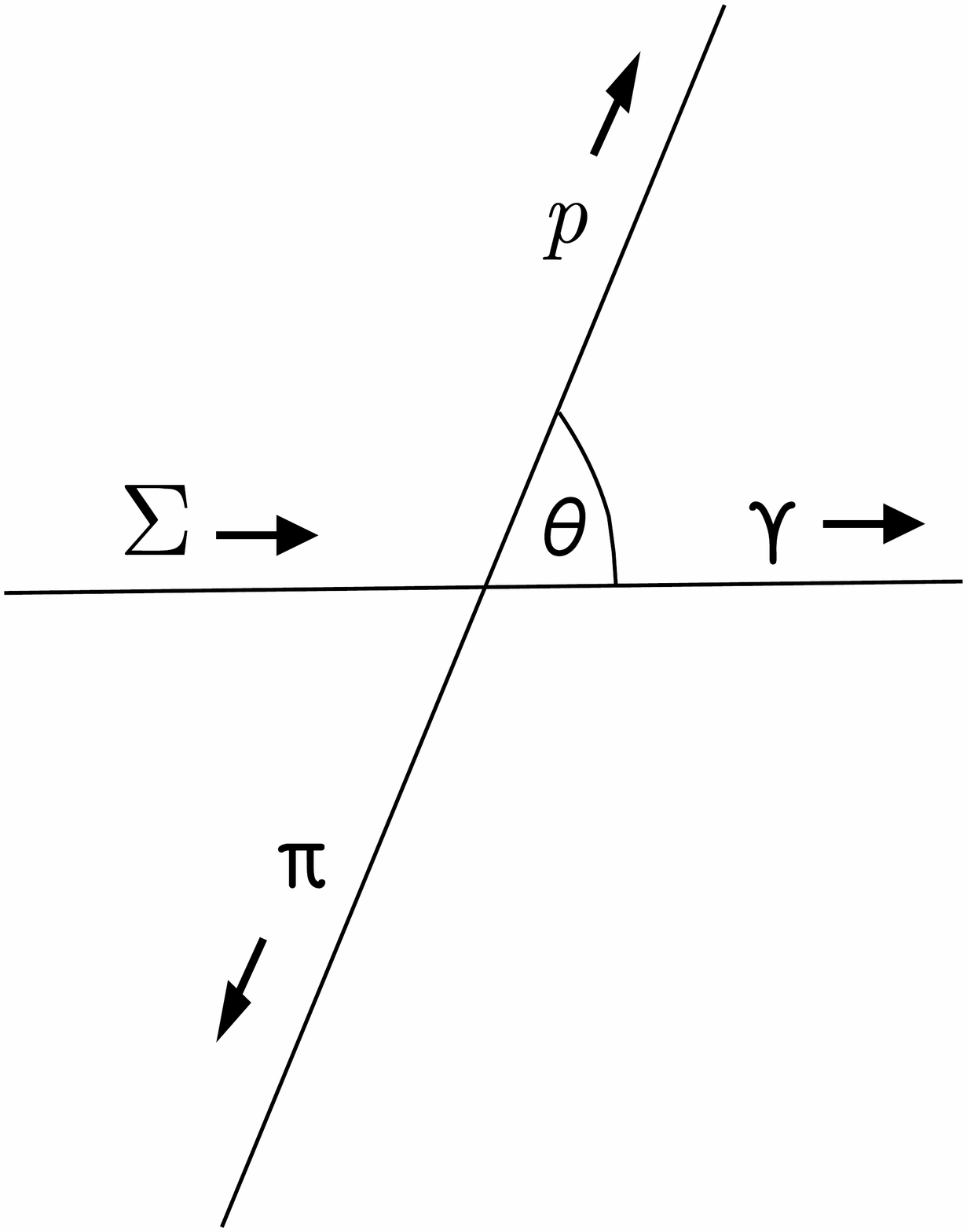}
  \end{minipage}  
  \caption{{\it Left-hand side:} Feynman diagram for the decay process $\Sigma^0 \to \gamma p \pi^-$. 
  {\it Right-hand side:} The decay process as seen in the $\Lambda$ rest frame.}
  \label{fig:figure1}
\end{figure}
For the full three-body decay $\Sigma^0 \to \gamma p \pi^-$, shown on the left-hand side of figure \ref{fig:figure1}, 
we start with the double-differential decay 
width \cite{Olive:2016xmw}
\begin{eqnarray}
  \frac{d^2\Gamma_{\Sigma^0 \to \gamma p \pi^-}}{dm_{23}^2 \, dm_{12}^2} 
  = \frac{1}{(2\pi)^3} \, \frac{1}{32 \,m_{\Sigma^0}^3} \, \left\langle \vert {\cal M}_3 \vert^2 \right\rangle \,,
  \label{eq:resgammadiff}
\end{eqnarray}
where we have introduced $m_{23}^2:=(p_\gamma+p_p)^2$ and $m_{12}^2:=(p_p+p_\pi)^2$. Our decay proceeds via the very narrow 
$\Lambda$ hyperon as an intermediate state. Consequently we will integrate over $m_{12}^2$ and focus on a single-differential 
decay width. In addition, we rewrite the $m_{23}^2$ dependence into a dependence on the angle $\theta$ between proton and photon 
defined in the rest frame of $\Lambda$. Note that in this frame we have $\vec p_\Sigma = \vec p_\gamma$ and 
$\vec p_p = - \vec p_\pi$ which defines a plane. This plane is depicted on the right-hand side of figure \ref{fig:figure1}.
One finds for the single-differential decay rate:
\begin{equation}
  \frac{d\Gamma_{\Sigma^0 \to \gamma p \pi^-}}{d\cos\theta} = 
  \frac{1}{(2\pi)^3} \, \frac{1}{64 \,m_{\Sigma^0}^3} \int dm_{12}^2 \, \left\langle \vert {\cal M}_3 \vert^2 \right\rangle \,
  \frac{m_{\Sigma^0}^2-m_{12}^2}{m_{12}^2} \, \lambda^{1/2}(m_{12}^2,m_p^2,m_\pi^2)  \,.
  \label{eq:gamdiff1}
\end{equation}

In \eqref{eq:resgammadiff}, \eqref{eq:gamdiff1} the matrix element ${\cal M}_3$ for the three-body decay is given by
\begin{eqnarray}
  \label{eq:matr3}
  {\cal M}_3 & = & \bar{u}_{p} \, (\mathcal{A}-\mathcal{B}\gamma_{5}) \, (\slashed{p}_\Lambda + m_\Lambda) \,
  (a \, \sigma_{\mu\nu} - b \, \sigma_{\mu\nu} \gamma_5 ) \, u_{\Sigma^0} \, (-i) p_\gamma^\nu \, \epsilon^{\mu *}  \, 
  D_\Lambda(m_{12}^2)  \nonumber \\
  & =: & {\cal M}_{\rm red} \; D_\Lambda(m_{12}^2)
\end{eqnarray}
with a $\Lambda$ resonance propagator
\begin{eqnarray}
  \label{eq:defD-res}
  D_\Lambda(s) := (s-m_\Lambda^2+ i m_\Lambda \Gamma_\Lambda)^{-1}
\end{eqnarray}
and the total width $\Gamma_\Lambda = 1/\tau_\Lambda$ where $\tau_\Lambda$ denotes the life time of the $\Lambda$.

Using the fact that the life time of the $\Lambda$ is extremely long \cite{Olive:2016xmw} one can carry out the integral 
that appears in \eqref{eq:gamdiff1}:
\begin{eqnarray}
  && \int dm_{12}^2 \, \left\langle \vert {\cal M}_3 \vert^2 \right\rangle  \,
  \frac{m_{\Sigma^0}^2-m_{12}^2}{m_{12}^2} \, \lambda^{1/2}(m_{12}^2,m_p^2,m_\pi^2) 
  \nonumber \\
  && = \int dm_{12}^2 \, \vert D_\Lambda(m_{12}^2) \vert^2 \, \left\langle \vert {\cal M}_{\rm red} \vert^2 \right\rangle  \,
  \frac{m_{\Sigma^0}^2-m_{12}^2}{m_{12}^2} \, \lambda^{1/2}(m_{12}^2,m_p^2,m_\pi^2) 
  \nonumber \\
  && = \frac{1}{m_\Lambda \Gamma_\Lambda} \, \int dm_{12}^2 \, 
  \frac{m_\Lambda \Gamma_\Lambda}{(m_{12}^2-m_\Lambda^2)^2+(m_\Lambda \Gamma_\Lambda)^2} \, 
  \left\langle \vert {\cal M}_{\rm red} \vert^2 \right\rangle  \nonumber \\ && \hspace*{12em} \times
  \frac{m_{\Sigma^0}^2-m_{12}^2}{m_{12}^2} \, \lambda^{1/2}(m_{12}^2,m_p^2,m_\pi^2) 
  \nonumber \\
  && \approx \frac{\pi}{m_\Lambda \Gamma_\Lambda} \, \left\langle \vert {\cal M}_{\rm red} \vert^2 \right\rangle  \,
  \frac{m_{\Sigma^0}^2-m_\Lambda^2}{m_\Lambda^2} \, \lambda^{1/2}(m_\Lambda^2,m_p^2,m_\pi^2) 
  \label{eq:doint}
\end{eqnarray}
where we have used
\begin{eqnarray}
  \label{eq:deltafuncrepr}
  \frac{m_\Lambda \Gamma_\Lambda}{(m_{12}^2-m_\Lambda^2)^2+(m_\Lambda \Gamma_\Lambda)^2} \approx \pi \delta(m_{12}^2-m_\Lambda^2) \,.
\end{eqnarray}
This replacement is completely appropriate when comparing to the experimental procedure where the intermediate $\Lambda$ is 
actually measured by a displaced vertex. 

Note that the reduced matrix element ${\cal M}_{\rm red}$ is now evaluated for onshell momenta of the $\Lambda$.
Then the average $\left\langle \vert {\cal M}_{\rm red} \vert^2 \right\rangle$ depends only on $m_{23}^2$ or, in other words, 
on the angle $\theta$ between proton and photon.
Finally one obtains for the single-differential decay rate:
\begin{eqnarray}
  \label{eq:finsddw}
  \frac{d\Gamma_{\Sigma^0 \to \gamma p \pi^-}}{d\cos\theta} = \frac12 \, \Gamma_{\Sigma^0 \to \gamma \Lambda} \, {\rm Br}_{\Lambda \to \pi^- p}
  \, (1-\alpha_{\Lambda}\alpha_{\Sigma^{0}}\cos\theta)
\end{eqnarray}
with the branching ratio ${\rm Br}_{\Lambda \to \pi^- p} := \Gamma_{\Lambda \to \pi^- p}/\Gamma_\Lambda$ and the asymmetries defined in
\eqref{eq:alphaSigma}, \eqref{eq:alphaLambda}. In terms of the number of events,
$N$, this reads
\begin{eqnarray}
  \label{eq:finsddwN}
  \frac{dN}{d\cos\theta} = \frac{N}{2}   \, (1-\alpha_{\Lambda}\alpha_{\Sigma^{0}}\cos\theta)   \,.
\end{eqnarray}

For the corresponding antiparticle decay chain one finds
\begin{eqnarray}
  \label{eq:finsddwNanti}
  \frac{d\bar N}{d\cos\theta} = \frac{\bar N}{2}   \, (1-\alpha_{\bar\Lambda}\alpha_{\bar\Sigma^{0}}\cos\theta)   \,.
\end{eqnarray}
with the number of events $\bar N$.

If one found an angular asymmetry in the particle and the antiparticle decays, then one could determine the parameters 
$\alpha_{\Sigma^{0}}$ and $\alpha_{\bar\Sigma^{0}}$ since the asymmetries $\alpha_{\Lambda}$ and $\alpha_{\bar\Lambda}$ have been 
measured \cite{Olive:2016xmw}. According to the definitions \eqref{eq:alphaSigma}, \eqref{eq:alphaSigmabar} and the 
relations given in the appendix an observable to test CP symmetry is given by
\begin{eqnarray}
  \label{eq:cp-test}
  {\cal O}_{\rm CP} := \alpha_{\Sigma^{0}} + \alpha_{\bar\Sigma^{0}}  \,,
\end{eqnarray}
i.e.\ this quantity vanishes if CP is conserved. Thus a non-vanishing value signals CP violation. A corresponding 
observable to test C symmetry is
\begin{eqnarray}
  \label{eq:c-test}
  {\cal O}_{\rm C} := \alpha_{\Sigma^{0}} - \alpha_{\bar\Sigma^{0}}  \,,
\end{eqnarray}

\section{Parameter estimates}
\label{sec:results}

We will now use three-flavor chiral perturbation theory \cite{Borasoy:2000pq,Ottnad:2009jw} to relate the 
photon-neutron-neutron couplings to the transition $\Sigma^0 \to \Lambda \gamma$. 
Given our moderate aim to establish an upper limit for the symmetry violating effects we will restrict ourselves 
to a next-to-leading-order (NLO) calculation. At this order the relevant terms of the effective Lagrangian have the form
\begin{eqnarray}
  \label{eq:NLO-rel}
  {\cal L}^{\rm NLO} = -\frac{3 e \kappa_n}{8 m_n} \, {\rm tr}(\bar B \sigma^{\mu\nu} \{Q,B\}) \, F_{\mu\nu}
  +\frac34 \, i \, d_n \, {\rm tr}(\bar B \sigma^{\mu\nu} \gamma_5 \{Q,B\}) \, F_{\mu\nu}   \,.
\end{eqnarray}
Here $F_{\mu\nu}$ is the electromagnetic field-strength tensor, $Q$ denotes the quark-charge matrix, 
and the baryons are collected in 
\begin{eqnarray}
  \label{eq:baroct}
  B = \left(
    \begin{array}{ccc}
      \frac{1}{\sqrt{2}}\, \Sigma^0 +\frac{1}{\sqrt{6}}\, \Lambda 
      & \Sigma^+ & p \\
      \Sigma^- & -\frac{1}{\sqrt{2}}\,\Sigma^0+\frac{1}{\sqrt{6}}\, \Lambda
      & n \\
      \Xi^- & \Xi^0 
      & -\frac{2}{\sqrt{6}}\, \Lambda
    \end{array}   
  \right)  \,.
\end{eqnarray}

Anticipating the consequences of this Lagrangian for the neutron properties we have already identified the low-energy constants
with the electromagnetic moments of the neutron. This will lead to predictions for the transition moments of $\Sigma^0$-$\Lambda$.
Note that the coefficient of the last term in \eqref{eq:NLO-rel} depends on the effective theta-vacuum 
angle \cite{Ottnad:2009jw}. The explicit dependence is 
irrelevant for the present purpose because we will only read off the NLO (i.e.\ tree-level) approximation to the 
neutron EDM and relate it on the one hand to the experimental upper limit \cite{Baker:2006ts} 
and on the other hand to the 
transition $\Sigma^0 \to \Lambda \gamma$. These identifications are correct at NLO accuracy. 

Keeping in \eqref{eq:NLO-rel} only the relevant terms of the baryon matrix \eqref{eq:baroct} yields
\begin{eqnarray}
  {\cal L}^{\rm NLO} & \to & \frac{e \kappa_n}{4 m_n} \, 
  \left(\bar n \sigma^{\mu\nu} n - \frac{\sqrt{3}}{2} \, \bar\Lambda \sigma^{\mu\nu} \Sigma^0 
    - \frac{\sqrt{3}}{2} \, \bar\Sigma^0 \sigma^{\mu\nu} \Lambda  \right) \, F_{\mu\nu}   \nonumber \\   && {}
  - \frac12 \, i d_n \, \left(\bar n \sigma^{\mu\nu} \gamma_5 n - \frac{\sqrt{3}}{2} \, \bar\Lambda \sigma^{\mu\nu} \gamma_5 \Sigma^0 
    - \frac{\sqrt{3}}{2} \, \bar\Sigma^0 \sigma^{\mu\nu} \gamma_5 \Lambda  \right) \, F_{\mu\nu}  \,.  \phantom{mj}
  \label{eq:NLO-conseq}  
\end{eqnarray}
Comparison with \eqref{eq:amp1} and recalling the decomposition \eqref{eq:decomdecFSI} leads to 
\begin{eqnarray}
  \label{eq:NLOpred-ab}
  a_D = -\frac{\sqrt{3} e \kappa_n }{4 m_n}  \,, \qquad  b_{D} = -\frac{\sqrt{3}}{2} \, i \, d_n  \,.
\end{eqnarray}

First we check the quality of our approximation by calculating the magnetic transition moment $\kappa_M$ 
given in \eqref{eq:defmagmoment} and \eqref{eq:abkappas}. We find
\begin{eqnarray}
  \label{eq:kappas-rel}
  \kappa_M = - \frac{\sqrt{3}}{2} \, \frac{m_{\Sigma^0}+m_\Lambda}{2 m_n} \, \kappa_n \approx 2.03
\end{eqnarray}
in reasonable agreement with the experimental value given in \eqref{eq:defmagmoment}. We conclude that the NLO relation 
\eqref{eq:NLOpred-ab} between the neutron EDM and the $\Sigma^0$-$\Lambda$ electric dipole transition 
moment is of sufficient quality for our estimates. 

Now we have all ingredients for an estimate of the asymmetry $\alpha_{\Sigma^0}$ defined in \eqref{eq:alphaSigma}. Using 
the second relation of \eqref{eq:NLOpred-ab} together with \eqref{eq:decomdecFSI} we obtain
\begin{eqnarray}
  \label{eq:asym-nEDM}
  \alpha_{\Sigma^0} \approx \frac{\sqrt{3} \, a \, d_n \sin\delta_F}{a^2+\frac34 d_n^2}   \,.
\end{eqnarray}
We utilize the current experimental upper bound of the neutron EDM \cite{Baker:2006ts}, 
$|d_{n}^{\rm{exp}}| \leq 2.9 \times 10^{-26} e\,\text{cm}$, together with the first relation of \eqref{eq:abkappas} and the 
experimental value of the magnetic transition moment \eqref{eq:defmagmoment}. This gives us the upper limit
\begin{eqnarray}
  \label{eq:upper-cons}
  \vert \alpha_{\Sigma^0} \vert \leq 3.0 \cdot 10^{-12}   \,.
\end{eqnarray}
Note that we have used here the most conservative ``limit'' on the phase caused by final-state 
interactions: $|\sin\delta_{F}| \leq 1$. Given that the final-state interaction between photon and $\Lambda$ is an 
electromagnetic effect, one would rather estimate  $|\sin\delta_{F}| \approx \alpha_{\rm QED}$ with the 
fine-structure constant $\alpha_{\rm QED}$. This would decrease the estimate \eqref{eq:upper-cons} by another two orders 
of magnitude. But since an upper limit should be conservative, we stick to \eqref{eq:upper-cons} which is anyway extremely small.

For the slope of the angular asymmetry in \eqref{eq:finsddwN} we deduce from \eqref{eq:upper-cons} an upper limit of 
\begin{eqnarray}
  \label{eq:upper-cons2}
  \vert \alpha_{\Lambda} \alpha_{\Sigma^0} \vert \leq 1.9 \cdot 10^{-12}
\end{eqnarray}
where the experimental value of the decay asymmetry for $\Lambda \to p \pi^{-}$ is given by $\alpha_{\Lambda} \approx 0.642$ 
\cite{Olive:2016xmw}. 

Finally we turn to the antiparticle decay chain. For the size of the corresponding asymmetry parameter we obtain 
\begin{eqnarray}
  \label{eq:asym-nEDM-anti}
  \alpha_{\bar\Sigma^0} \approx \frac{\sqrt{3} \, a \, d_n \sin\delta_F}{a^2+\frac34 d_n^2} 
\end{eqnarray}
i.e.\ the very same estimate as in \eqref{eq:asym-nEDM}. This is not surprising given the 
fact that the QCD theta-vacuum-angle term conserves C and violates CP; see also the discussion in the appendix. 
Thus it is useful to provide a corresponding upper limit for the CP-test 
observable \eqref{eq:cp-test}:
\begin{eqnarray}
  \label{eq:upper-cons2cp}
  \vert {\cal O}_{\rm CP} \vert \leq 6.0 \cdot 10^{-12}  \,.
\end{eqnarray}

\section{Summary and Outlook}
\label{sec:summary}

We have provided a framework to search for P and CP violation in the decay $\Sigma^0 \to \Lambda \gamma$ 
utilizing the subsequent weak decay of the $\Lambda$ to achieve an angular asymmetry. The driving term is an electric transition
dipole moment. For our concrete estimates we have 
used the CP violating QCD theta-vacuum-angle term. Since this term gives also rise to an electric dipole moment of the neutron 
one can use chiral perturbation theory to relate the two electric dipole moments. Using the experimental upper limit for 
the neutron electric dipole moment provides upper limits for the angular asymmetry in our $\Sigma^0$ decay and also for 
the parameter that tests CP violation by comparing the angular asymmetries for particle and antiparticle. We have found tiny
effects which implies that any experimental significance would point to physics beyond QCD, even if extended 
by a theta-vacuum-angle term.

In principle, also P violation without or with little CP violation can lead to an angular asymmetry. This can be caused for 
instance by the weak theory. An estimate of this effect is beyond the scope of the present paper. 
Yet we would like to stress that the CP violation in the weak theory is 
very small. Thus an observation beyond the tiny effect \eqref{eq:upper-cons2cp} of a CP violating angular asymmetry 
as deduced from the $\Sigma^0$ and $\bar \Sigma^0$ decays
would point to physics beyond the standard model.

\appendix
\section{Phases and discrete symmetries}
\label{sec:appendix}

An effective Lagrangian for the $\Sigma^{0}$-$\Lambda$ transition takes the form
\begin{eqnarray}
  \mathcal{L}_{\Sigma^{0} \Lambda} & = & \frac12 \, a_{D} \, \bar \Lambda \sigma_{\mu\nu} \Sigma^{0} \, F^{\mu\nu} 
  + \frac12 \, \bar a_{D} \, \bar \Sigma^{0} \sigma_{\mu\nu} \Lambda \, F^{\mu\nu}   \nonumber \\ && {}
  - \frac12 \, b_{D} \, \bar \Lambda \sigma_{\mu\nu} \gamma_{5} \Sigma^{0} \, F^{\mu\nu} 
  - \frac12 \, \bar b_{D} \, \bar \Sigma^{0} \sigma_{\mu\nu} \gamma_{5} \Lambda \, F^{\mu\nu}   \,.
  \label{eq:interactionlagrangian}
\end{eqnarray}
Hermiticity requires $\bar{b}_{D}= -b_{D}^{*}$ and $\bar a_D = a_D^*$. We still have the freedom to redefine the field 
$\Sigma^0$ by an arbitrary phase. We choose this phase such that $a_D = \bar a_D$ is positive and real. 

Using the transformation properties of fermion bilinears \cite{pesschr} it is easy to show that the $a$ terms in our interaction 
Lagrangian (\ref{eq:interactionlagrangian}) conserve P
and C symmetry while the $b$ terms break P. In addition, 
C symmetry implies $b_{D}=\bar{b}_{D}$ (i.e.\ $b_{D}$ is purely imaginary). Note that conserved C and broken P implies that CP is
broken. This is the breaking pattern caused by the QCD theta-vacuum-angle term, as one can also read off 
from \eqref{eq:NLOpred-ab}. 
What about the case when CP is conserved? This is fulfilled 
when $b_{D}=-\bar{b}_{D}$ (i.e.\ $b_{D}$ is purely real) \cite{Dreitlein:1962zz}. 
For a P violating decay, conservation of CP implies that C is violated. Note that for the case when P, C and CP are violated 
$b_{D}$ is neither purely real nor purely imaginary.

Finally we recall that electromagnetic final-state interactions induce a phase shift $\delta_{F}$ that is the same for particle
and antiparticle. Thus
\begin{eqnarray}
b = b_{D} \, e^{i\delta_{F}} \,, \qquad  \bar{b}= \bar{b}_{D} \, e^{i\delta_{F}}.
\end{eqnarray}

\bibliography{lit}{}
\bibliographystyle{elsarticle-num}
\end{document}